\documentclass[%
 reprint,
 amsmath,amssymb,
 aps,
]{revtex4-1}

\usepackage{geometry}
\usepackage{color}
\usepackage[at]{easylist}
\usepackage{tikz}
\usetikzlibrary{positioning}
\usetikzlibrary{decorations.pathreplacing}
\usepackage{qcircuit}
\usepackage{braket}
\usepackage{lcg} 
\reinitrand[counter=random,first=10,last=99,seed=0]
\usepackage{ifthen}

\newcommand{\argmax}{\mathop\textrm{ argmax}\limits}

\begin{document}

\title{	
A Quantum Algorithm for Finding $k$-Minima
}

\author{Kohei Miyamoto}
 \email{miyamotokohei@protonmail.com}
\author{Masakazu Iwamura}%
 \email{masa@cs.osakafu-u.ac.jp}
\author{Koichi Kise}
 \email{kise@cs.osakafu-u.ac.jp}
\affiliation{
 Department of Computer Science and Intelligent Systems, 
Graduate School of Engineering, 
Osaka Prefecture University
}%

\date{\today}


\begin{abstract}
We propose a new \textit{finding $k$-minima} algorithm and prove that its query complexity is $\mathcal{O}(\sqrt{kN})$, where $N$ is the number of data indices.
Though the complexity is equivalent to that of an existing method, the proposed is simpler.
The main idea of the proposed algorithm is to search a good threshold that is near the $k$-th smallest data.
Then, by using the generalization of amplitude amplification, all $k$ data are found out of order and the query complexity is $\mathcal{O}(\sqrt{kN})$.
This generalization of amplitude amplification is also not well discussed and we briefly prove the query complexity.
Our algorithm can be directly adapted to distance-related problems like $k$-nearest neighbor search and clustering and classification.
There are few quantum algorithms that return multiple answers and they are not well discussed.
\end{abstract}

\pacs{Valid PACS appear here}
                              
\maketitle

\section{Introduction}        


We propose an $\mathcal{O}(\sqrt{kN})$ quantum algorithm for finding $k$-minima, where $N$ is the number of data indices.
Our algorithm finds the $k$ smallest from $N$ data indices.
D\"urr and H{\o}yer have originally proposed an $\mathcal{O}(\sqrt{N})$ quantum algorithm for finding one minimum~\cite{durr1996quantum}.
Then, D\"urr, et al. have proposed an $\mathcal{O}(\sqrt{kN})$ quantum algorithm 
for finding $k$-minima~\cite{durr2006quantum}.
Though the proposed algorithm has the same query complexity as the one proposed by D\"urr and Heiligman~\cite{durr2006quantum}, the proposed is simpler.
One reason is that they take into account types of data.
They want to use this algorithm as a part of graph algorithm,
therefore their algorithm is not purely designed as finding $k$-minima algorithm.

The proposed algorithm is based on following three algorithms: finding minimum (FM) algorithm~\cite{durr1996quantum}, quantum counting (QC) algorithm~\cite{brassard1998quantum,brassard2002quantum} and amplitude amplification (AA)~\cite{boyer1996tight,brassard2002quantum}.
FM algorithm and QC algorithm are used to find a good threshold index, and AA is used to find all $k$ indices whose values are less than the value of the threshold index.

In addition, this paper contributes to the following two.
First, we explicitly distinguish gate complexity and query complexity by defining new symbols.
Second, we re-formulate FM algorithm and \textit{finding $k$-minima} algorithm following the manner of AA.
Therefore, all of them can be compared more easily and clearly.
AA is a quantum database search algorithm and is a generalization of Grover's algorithm~\cite{grover1996fast}.
From $N$ indices, AA searches one of $k$ indices that satisfy some certain condition with the query complexity of $\mathcal{O}(\sqrt{N/k})$.

Many quantum algorithms do not directly return multiple answers because the measurement of quantum states collapses the state of superposition.
Therefore, many trials are required to obtain all the results.
This is the disadvantage of quantum algorithms and such trials sometimes increase linearly for the number of results.
Hence, $\mathcal{O}(k)$ trials are required for $k$ results.
However, our algorithm solves the problem that returns $k$ results with $\mathcal{O}(\sqrt{k})$.
We use a generalization of AA that searches all $k$ indices from $N$ indices.
We call this algorithm \textit{searching all marked $k$-indices} algorithm.
This problem is solved in $\mathcal{O}(\sqrt{kN})$ query complexity~\cite{ambainis2004quantum,ambainis2005new,klauck2007quantum,dorn2010note,aimeur2013quantum}.

\textit{Finding $k$-minima} algorithm can be applied to $k$-nearest neighbor search like \cite{wiebe2014quantum} and other quantum machine learning methods such as clustering and classification~\cite{aimeur2007quantum,aimeur2013quantum,lloyd2013quantum,schuld2015introduction}.

\section{Complexity Measures \label{sec:complexity}}

Two kinds of complexity measures are used in quantum algorithms. 
One is \textit{quantum gate complexity} that is based on the number of quantum gates to solve a problem.
The other is \textit{query complexity} that is based on the number of queries to solve a problem.
For these two complexity measures, the same mathematical expression $\mathcal{O}$ is used.
However, in this paper, we explicitly distinguish them by using $\mathcal{O}_g$ for the quantum gate complexity and $\mathcal{O}_q$ for the query complexity.
%
For example, making desired quantum states of a $d$ dimensional vector in $n=\log{d}$ [qubit] requires $\mathcal{O}_g (n)=\mathcal{O}_g(\log{d})$~\cite{grover2002creating,kaye2004quantum,soklakov2006efficient,lloyd2013quantum}, quantum fourier transform of $n$ [qubit] is $\mathcal{O}_g(n\log{N})$~\cite{hales2000improved} and AA's complexity is $\mathcal{O}_q(\sqrt{N/k})$~\cite{boyer1996tight,brassard2002quantum}.

\subsection{Quantum Gate Complexity}

Quantum algorithms based on quantum gates solve the problems by arranging the quantum gates.
The Hadamard gate and controlled-NOT gate are the most basic quantum gates.
It is known that any kinds of unitary transformation can be approximated by these two gates.
This means that the quantum gate complexity can be measured by the number of these two gates.

\subsection{Query Complexity}

While the quantum gate complexity is easy to understand, most of the quantum algorithms are evaluated based on the query complexity.
This is because many quantum algorithms use oracles.

A query is input to the oracle and query complexity is defined by the number of queries to get an answer.
In other words, query complexity measures how many times oracles are called.
Oracle is usually treated as a single quantum gate.
Therefore, oracles are sometimes called black-box and query model is called black-box model.

To analyze the query complexity, some kinds of methods are proposed.
For example, polynomial methods~\cite{beals2001quantum,nielsen2002quantum} and adversary methods~\cite{ambainis2002quantum,spalek2004all}.
Polynomial methods generalize a boolean function $f(x):\{0,1\}^n \rightarrow \{0,1\}$ to a polynomial function $p(x):\mathbb{R}^N \rightarrow \mathbb{R}$.
By analyzing the degree of polynomial function $p(x)$, lower bounds can be known.
Strong direct product theorem is an important theorem to measure the complexity in both polynomial and adversary methods~\cite{klauck2007quantum,ambainis2005new,ambainis2009new}.

\section{Preliminaries}

In this section, we define mathematical symbols and complexity while presenting how oracle $f$ is defined in AA.
AA is one of the most important quantum algorithms and we present FA algorithm following the manner of AA.

All kinds of data are treated as a binary representation in a digital computer.
However, quantum computer not only treats binary representation with superposition but also treats analog representation by superposition of quantum bits~\cite{mitarai2019quantum}.
Digital representation can have multiple numbers in one time by superposition and analog representation can have one vector that the square of the number is represented by a probability.
We will not describe more detail about the representation in this paper.

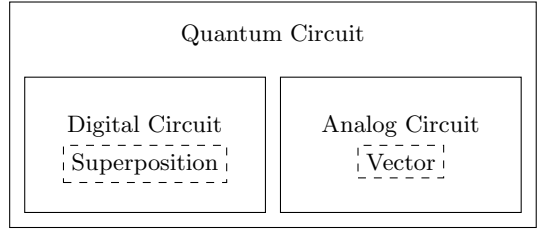
\begin{figure}[t]
	\centering
\begin{tikzpicture}
	\newcommand{\x}{0.2}
	\draw[] (0,0) rectangle ++(7,3) node[midway, above=8mm]{Quantum Circuit};
	\draw[] (\x,\x) rectangle (7/2-\x/2,2) node[midway,above]{Digital Circuit} node[midway,below,draw,rectangle,dashed]{Superposition};
	\draw[] (7/2+\x/2,\x) rectangle (7-\x,2) node[midway,above]{Analog Circuit} node[midway,below,draw,rectangle,dashed]{Vector};
\end{tikzpicture}
	\caption{Quantum circuit is composed of a digital circuit and an analog circuit. 
	Digital circuit uses superposition of binary representation. 
	Analog circuit uses the vector of the number such that the square of the number becomes a probability.}
	\label{fig:quantumanddigital}
\end{figure}

An output that we can measure is binary representation in a quantum computer.
Therefore, the last output of the algorithm is binary representation.
However, we can use either way in the middle of the quantum circuit.
We call circuit that uses binary representation as a digital circuit and algorithm that use analog representation as an analog circuit.
This means that the quantum circuit contains a digital circuit and an analog circuit like Figure~\ref{fig:quantumanddigital}.
We have to mention that the input and output of this digital circuit can be superposition in this paper.
Many quantum circuits use both a digital circuit and an analog circuit.
Moreover, quantum bits sometimes go back and force digital representation and analog representation.

We use AA as a base of our explanation.
AA searches one index that satisfies a condition.
The condition is given by oracle.
To use AA, we have to define oracle that fits a problem.

Let $D$ be a set of indices and each of which is tied to value.
$|D|=N$ and $x \in D$ is represented by binary.
\[
	x=\{0,1\}^{\log{N}}
\]
For simplicity, 
\[
	N=|D|=2^n \text{, where } n \in \mathcal{N}
\]
We want to find index $x \in D$ that satisfies a condition.
For example, index $x$ that has less value than the threshold value.

Let $f(x)$ be a boolean function such that
\[ 
f(x) : \{0,1\}^{\log{N}} \rightarrow \{0,1\}. 
\]
If $f(x)=1$, $x$ satisfies a condition $f$ and if $f(x)=0$, $x$ does not satisfy a condition $f$.
\[
	f(x)=
	\begin{cases}
		1, &
		\text{if }
		x \text{ satisfies condition $f$,}
		\\
		0, &
		\text{if }
		x \text{ does not satisfy condition $f$.}
	\end{cases}
\]
In some paper, $x \in D$ is called marked index if $f(x)=1$,
and $x \in D$ is called unmarked index if $f(x)=0$.

\begin{figure}[t]
	\input{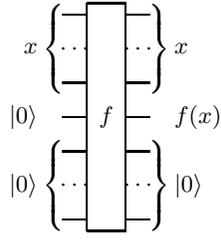}
	\caption{A quantum circuit that converts index $x$ to $f(x)$.
		$x$ is a binary index and $f(x)$ is a binary.
		The bottom input $\ket{0}$ and output $\ket{0}$ is a workspace for quantum computing. 
		Most of the oracles use workspaces like this, however, it is often omitted in a quantum circuit.
		The output other than the desired output must be the same condition as the input condition.
		In this case, only the output $f(x)$ is different from the input
		and others are the same.
		This is because the measurement of non-desired output should not influence on the desired output.
		To avoid this, we have to add a gate to inverse the process.
		However, we omit such gate for simplicity in this paper.
	}
	\label{fig:oraclecircuit}
\end{figure}

This kind of boolean function is called oracle and thought as a single quantum gate.
If we express $f$ in quantum circuit, it becomes Figure~\ref{fig:oraclecircuit}.
An input $x$ is a binary representation and $x$ is superposed and input to the quantum circuit $f$ in AA.

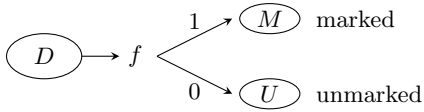
\begin{figure}[t]
	\centering
\begin{tikzpicture}
\draw(0,0) circle[x radius=5mm, y radius=3mm] node[]{$D$};
\draw[-stealth](0.5,0)--(1,0) node[right]{$f$};
\draw[-stealth](1.5,0)--(2.5,0.5) node[midway,above]{$1$};
\draw[-stealth](1.5,0)--(2.5,-0.5) node[midway,below]{$0$};
\draw(3,0.5) circle[x radius=4mm, y radius=2mm] node[]{$M$} node[right=5mm]{marked};
\draw(3,-0.5) circle[x radius=4mm, y radius=2mm] node[]{$U$} node[right=5mm]{unmarked};
\end{tikzpicture}
	\caption{$D$ is divided into $M$ and $U$ by $f$}
	\label{fig:DMUf}
\end{figure}

Thinking $f$ from another point of view, $f$ divide $D$ into two sets.
Let $M$ and $U$ be sets of indices such that $f(x)=1$ (marked) and $f(x)=0$ (unmarked) respectively.
\begin{align*}
	M&=\lbrace
	x \mid x \in D, f(x)=1
	\rbrace
	\\
	U&=\lbrace
	x \mid x \in D, f(x)=0
	\rbrace
\end{align*}
$M \cup U = D$ and $M\cap U = \phi$.
$D$ is divided into $M$ and $U$ like Figure~\ref{fig:DMUf}.

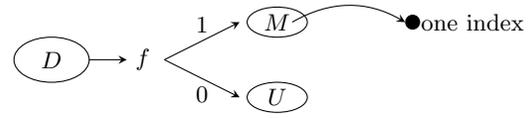
\begin{figure}[t]
	\centering
\begin{tikzpicture}
\draw(0,0) circle[x radius=5mm, y radius=3mm] node[]{$D$};
\draw[-stealth](0.5,0)--(1,0) node[right]{$f$};
\draw[-stealth](1.5,0)--(2.5,0.5) node[midway,above]{$1$};
\draw[-stealth](1.5,0)--(2.5,-0.5) node[midway,below]{$0$};
\draw(3,0.5) circle[x radius=4mm, y radius=2mm] node[]{$M$} node[right=5mm]{};
\draw[-stealth](3.2,0.5) to[bend left] ++(1.5,0);
\fill(4.8,0.5) circle[radius=1mm] node[right]{ one index };
\draw(3,-0.5) circle[x radius=4mm, y radius=2mm] node[]{$U$} node[right=5mm]{};
\end{tikzpicture}
	\label{fig:AA}
	\caption{Overview of AA. 
	AA finds one index from $M$ and $M$ is a subset of $D$. $M= \lbrace x \mid x \in D, f(x)=1 \rbrace$.}
\end{figure}

Original AA finds one index $x \in D$ such that $f(x)=1$.
In other words, find $x$ from $M$.
We have to mention that AA does not find all $x\in M$.
AA can find only one index in $M$ like figure~\ref{fig:AA}.

When we want to find all $M$ from $D$, we have to update $M$ each time $x \in M$ found.
Let $|M|=k$ and all $x \in M$ are found the order of $\lbrace x_1,...,x_k \rbrace$.
In this case, AA is called $k$ times and $x_i$ means that $x_i \in M$ is found at $i$ times calling of AA.
In general, we cannot know the order of finding before applying.
In each step, $M$ is updated and one index that found is removed.
Let $M_i$ be the $i$ iteration of $M$ and $M_1 = M$.
\begin{align*}
	M_i &= M_{i-1}\backslash x_{i-1} \\
		&= M_1 \backslash \lbrace x_1 ,..., x_{i-1} \rbrace \\
		&= \lbrace x_i ,..., x_k \rbrace \\
	|M_i| &= k-i-1
\end{align*}
See appendix for more detail about finding all marked $k$-indices.

Finally, we present how to design $f$ by showing one example.
AA needs oracle $f$ however it is not given in algorithm and only input and output are assigned.
For example, if we want to find one of the indices that has less value than the value of threshold index $t$.

\begin{figure}[t]
	\input{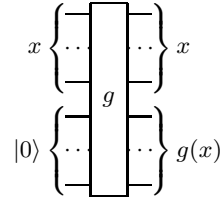}
	\caption{A quantum circuit $G$ that converts index $x$ to value $g(x)$.
		$x$ is a binary index.
		The input index and output index must be the same condition.
		This is because measurement of index should not influence on the output value $g(x)$.
	}
	\label{fig:indexdatacircuit}
\end{figure}

Let $f_t$ be a required oracle to solve this problem on AA and
let $g$ be a function that returns a value by which index $x \in D$.
In other words $g(x)$ returns a value that is combined to index $x \in D$.
$g$ is also oracle and the quantum circuit is like Figure~\ref{fig:indexdatacircuit}.
The gate complexity of this oracle $g$ is $\mathcal{O}_g(\log d)$ as we described in Section~\ref{sec:complexity}. 
$d$ is a dimension of $g(x)$.
In other words, $d$ is a number of quantum bit.
For more detail, see~\cite{grover2002creating,kaye2004quantum,soklakov2006efficient,lloyd2013quantum}.

The input $x$ and output $g(x)$ of quantum circuit $g$ is also binary representation and can be superposed. 
Therefore output can be superposed if input index $x$ is superposed.
Figure~\ref{fig:indexdata} is an example of how $g$ is working.
\begin{figure}[t]
	\centering
\begin{tikzpicture}
\draw(0.25,1.5) node[]{Indices} ++(2.5,0) node[]{Values};
\draw(0.25,1) node[]{$x$} ++(2.5,0) node[]{$g(x)$};
\foreach \i in {0,1,2, ..., 6}{
	\draw(0,-\i/2) rectangle (0.5,-\i/2+0.5);
	\draw(0.25,-\i/2+0.25) node[]{$\i$};
}
\foreach \i in {0,1,2, ..., 6}{
	\rand
		\draw(2,-\i/2) rectangle ++(2,0.5);
	\draw(3,-\i/2+0.25) node[]{$g(\i)=\arabic{random}$};
}
\node[] at (0.25, -7/2+0.25) {$\vdots$};
\node[] at (3, -7/2+0.25) {$\vdots$};
\draw[-stealth](1,-1.25)-- ++(0.5,0) node[midway,above]{$g$};
\end{tikzpicture}
	\caption{Example of the input $x$ and output $g(x)$ of oracle $g$.
	}
	\label{fig:indexdata}
\end{figure}
Then, the oracle $f$ is expressed like this.
\begin{align}
	f_t(x)
	=
	\begin{cases}
		1, & \text{if } g(x)<g(t),\\
		0, & \text{if } g(x) \ge g(t).
	\end{cases}
\end{align}
Here, $t$ is a threshold index in $D$.
We use this oracle $f_t$ in section~\ref{sec:fm} for FM algorithm.
Figure~\ref{fig:exampleoff} shows an example of how $f$ is working.
\begin{figure}[t]
	\centering
\begin{tikzpicture}
\draw(0.25,1) node[]{$x \in D$} ++(3,0) node[]{$g(x)$} ++(1.5,0) node[] {$g(t)$} ++(1.5,0) node[]{$f_t(x)$};
\foreach \i in {0,1,2, ..., 6}{
	\draw(0,-\i/2) rectangle (0.5,-\i/2+0.5);
	\draw(0.25,-\i/2+0.25) node[]{$\i$};
}
\node[] at (0.25,-7/2+0.25) {$\vdots$};
\draw(0,-2+0.25) node[left]{$t \rightarrow$};
\foreach \i in {0,1,2, ..., 6}{
	\rand
		\ifthenelse{ \i = 4 }{
			\draw(2,-\i/2) rectangle ++(2,0.5) ++(-1, -0.25) node[]{$g(\i)=55$};
		}{
			\draw(2,-\i/2) rectangle ++(2,0.5) ++(-1, -0.25) node[]{$g(\i)=\arabic{random}$};
		}

	\ifthenelse{ \i = 4 }{
		\draw(5.5,-\i/2+0.25) node[]{$= 55 \rightarrow f_t(\i)=0$};
	}{
		\ifthenelse{ \arabic{random} < 55}{
			\draw(5.5,-\i/2+0.25) node[]{$< 55 \rightarrow f_t(\i)=1$};
		}{
			\draw(5.5,-\i/2+0.25) node[]{$\ge 55 \rightarrow f_t(\i)=0$};
		}
	}
}
\draw[-stealth](1,-1.25)-- ++(0.5,0) node[midway,above]{$g$};
\node[] at (3,-7/2+0.25) {$\vdots$};
\node[] at (4.75,-7/2+0.25) {$\vdots$};
\node[] at (6,-7/2+0.25) {$\vdots$};
\end{tikzpicture}
	\caption{Example of indices, value and $f_t$. When threshold $t$ is $4$.
		Superposed indices $x$ are converted to values $g(x)$ and compared by the threshold value $g(t)$.
	}
	\label{fig:exampleoff}
\end{figure}

Let $\text{CMP}$ be a function that compare two values and return $1$ or $0$.
\begin{align*}
	\text{CMP}(v_1, v_2) =
	\begin{cases}
		1, & \text{if } v_1 < v_2,\\
		0, & \text{if } v_1 \ge v_2.
	\end{cases}
\end{align*}
The computational complexity of oracle $\text{CMP}$ is $\mathcal{O}_g(\log{d})$, 
because the input $v_1$ and $v_2$ are binary representation.

\begin{figure}[t]
	\input{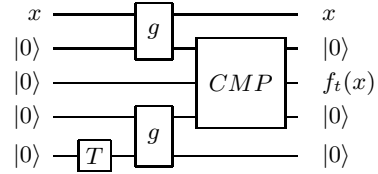}
	\caption{Quantum circuit of oracle $f_t$.
		The index $t$ is created by the gate $T$ in the circuit.
		In this example, $\text{CMP}$ gate is symmetry.
		The center input is for output $f_t(x)$ and
		above and below inputs are for two values.
		If above input is less than the input of below,
		$\text{CMP}$ returns $0$. Otherwise, returns $1$.
	}
	\label{fig:qcircuitft}
\end{figure}

From the above, a quantum circuit of $f_t$ is expressed as Figure~\ref{fig:qcircuitft}.
In this circuit, the upper input $x$ and the middle output $f_t(x)$ are used in AA and other inputs and outputs are not used.
However, we have to keep unchanged of inputs and outputs so as not to influence the states of $f_t(x)$.

\section{Finding Minimum Algorithm\label{sec:fm}}

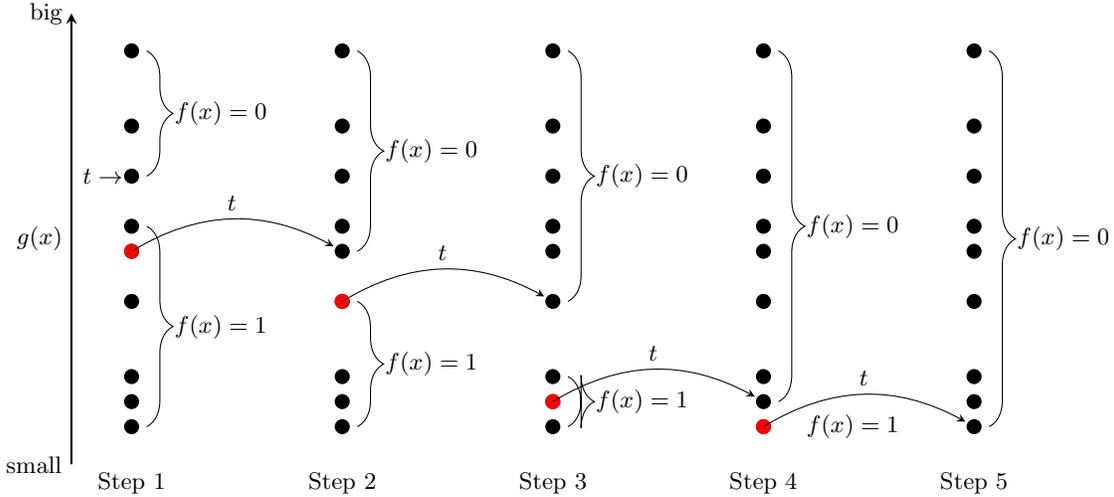
\begin{figure*}[t]
	\centering
\begin{tikzpicture}
\draw[-stealth,thick](2,-0.5) node[left]{small}--++(0,6) node[midway,left]{$g(x)$} node[left]{big};
\newcommand{\x}[1]{#1*2.8}
\draw(\x{1},10/3) node[left]{$t\rightarrow$};
\foreach \i in {1,2,3,4,5}{
	\node(O) at (\x{\i},0){};
	\draw[](O)+(0,-0.5) node[below]{Step \i};
	\foreach \j  in {0,1,2,5,7,8,10,12,15}{
		\node[] (n) at (\x{\i},\j/3){};
		\fill(n) circle[radius=1mm];
		\ifthenelse{ 
			\(\i=1 \AND \j=10\)  \OR
				\(\i=2 \AND \j=7\)  \OR
				\(\i=3 \AND \j=5\)  \OR
				\(\i=4 \AND \j=1\)  \OR
				\(\i=5 \AND \j=0\)
		}{
			\ifthenelse{ \i > 1 }{
				\fill[red](n)+(-2.8,0) circle[radius=1mm];
				\draw[-stealth](n) ++(-2.8,0) to[bend left=30] node[midway, above]{$t$} (n);
			}{}
			\draw [decorate,decoration={brace,mirror,amplitude=10pt}] (n)+(0.2,0) -- (\x{\i}+0.2,5) node [midway,right=0.8em] {$f(x)=0$};
		}{}
		\ifthenelse{ 
			\(\i=1 \AND \j=8\)  \OR
				\(\i=2 \AND \j=5\)  \OR
				\(\i=3 \AND \j=2\)  \OR
				\(\i=4 \AND \j=0\)
		}{
			\draw [decorate,decoration={brace,amplitude=10pt}] (n)+(0.2,0) -- (\x{\i}+0.2,0) node [midway,right=0.8em] {$f(x)=1$};
		}{}
	}
}
\end{tikzpicture}
	\caption{The updating process of threshold $t$ in FM algorithm.
	The vertical axis represents the value of indices.
		When the first threshold $t$ is selected, all points are divided into two.
		One is $f(x)=0$ and the other is $f(x)=1$.
		AA randomly selects one of the points such that $f(x)=1$.
		The selected point, drawn in red, is used as a new threshold index $t$ in the next step.
		The number of points satisfying $f(x)=1$ decreases as the step goes.
		Finally (at the Step 5 in this example), the algorithm ends as no points satisfy $f(x)=1$.
	}
	\label{fig:fmprocess}
\end{figure*}

We begin with presenting an overview of FM algorithm
because it is a simple application of AA and helps understand the succeeding algorithms.
An example of the procedure of the algorithm is shown in Figure~\ref{fig:fmprocess}.
This algorithm finds the minimum from $D$ with the complexity of $\mathcal{O}_q(\sqrt{N})$
by iteratively updating threshold index $t$ by oracle $f_t$.
Oracle $f_t(x)$ is a function that marks index $x$ such that $g(x)<g(t)$.
That is,
\begin{align}
    \label{eqn:FM_oracle}
	f_t(x)
	=
	\begin{cases}
		1, & \textrm{if } g(x)<g(t),\\
 		0, & \textrm{if } g(x) \ge g(t).
	\end{cases}
\end{align}
The oracle marks the indices that have smaller values than the value of threshold $t$.
Setting a marked index as the new threshold, the value of threshold decreases.
Eventually, we obtain the index that has the minimum value.

In summary, FM algorithm is given as follows.
\begin{enumerate}
\item Select threshold index $t$ from $D$ uniformly at random.
\item Repeat the following process more than 
	\[
		22.5\sqrt{N}+1.4\log^2{N}
	\]
	times.
	\begin{enumerate}
		\item Find index $x$ such that $f_t(x)=1$.
		\item Set the found index $x$ as the threshold index $t$.
	\end{enumerate}
\item Return $t$ as the index that has the minimum value in $D$.
\end{enumerate}

\section{Conventional Finding $k$-Minima Algorithm}

D\"urr and H{\o}yer have proposed an $\mathcal{O}_q(\sqrt{kN})$ algorithm that finds the $k$ smallest indices \textit{of different types} from $D$~\cite{durr2006quantum}.
As the algorithm considers \textit{types}, it is complex.
Hence, we present it in an easier-to-understand way.

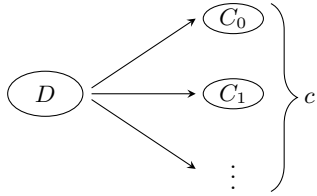
\begin{figure}[t]
	\centering
\begin{tikzpicture}
\draw(0,0) node[](o){} circle[x radius=5mm, y radius=3mm] node[]{$D$};
\node[](d) at (0.5,0){};
\draw[-stealth](d) -- (2,1) node[](c0){};
\draw[-stealth](d) -- (2,0) node[](c1){};
\draw[-stealth](d) -- (2,-1)node[](c2){};
\draw(c0)+(0.5,0) circle[x radius=4mm, y radius=2mm] node[]{$C_0$};
\draw(c1)+(0.5,0) circle[x radius=4mm, y radius=2mm] node[]{$C_1$};
\draw(c2)+(0.5,0) node {$\vdots$};
\draw [decorate,decoration={brace,amplitude=10pt}] (3,1.2) -- ++(0,-2.5) node [midway,right=1em] {$c$};
\end{tikzpicture}
	\caption{$D$ is divided into $C_i$ by types.}
	\label{fig:newcondition}
\end{figure}

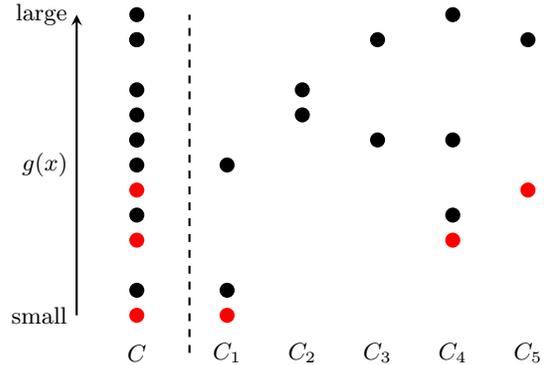
\begin{figure}[tb]
	\centering
\begin{tikzpicture}
	\draw[-stealth,thick](-1,0) node[left]{small}--(-1,4) node[midway,left]{$g(x)$} node[left]{large};
	\draw[dashed,thick](0.5,-0.5)--++(0,4.5);
	\foreach \j / \i  in {
		0/1,1/1,
		3/4,4/4,5/5,6/1,7/3,7/4,8/2,9/2,11/3,11/5,12/4
		}{
		\node[] (n) at (\i,\j/3){};
		\node[] (c) at (-0.2,\j/3){};

		\ifthenelse{ \j = 0 \OR \j = 3 \OR \j = 5 }{
			\fill[red](c) circle[radius=1mm];
			}{
			\fill(c) circle[radius=1mm];
		}
		\ifthenelse{ 
			\(\i = 1 \AND \j= 0 \) \OR
			\( \i = 4 \AND \j = 3 \) \OR
			\( \i = 5 \AND \j = 5 \)
			}{ 
			\fill[red](n) circle[radius=1mm];
			}{
			\fill(n) circle[radius=1mm];
		}
	}
	\node[] at (-0.2,-0.5) {$C$};
	\foreach \i in { 1,2,...,5 }{
		\node[] at (\i,-0.5) {$C_\i$};
	}
\end{tikzpicture}
	\caption{Overview of D\"urr and H{\o}yer's \textit{finding $k$-minima} algorithm with different types in the case of $c=5$ and $k=3$.
		$C$ is the set of all values of $D$.
		$C_i$ is the set of values that belong to type $i$.
		Each $C_i$ can contain at most one minimum.
		This is equal to the case that all of $k$ indices are different types.
	}
	\label{fig:types}
\end{figure}

\begin{figure*}[tb]
	\centering
\begin{tikzpicture}
	\draw[-stealth,thick](-1,0) node[left]{small}--(-1,4) node[midway,left]{$g(x)$} node[left]{large};
	\draw[dashed,thick](0.2,-0.5)--++(0,4.5);
	\foreach \j / \i  in {
		0/1,
		1/2,
		6/3,
		8/4,
		9/6,
		7/7,
		11/8,
		3/9,
		4/10,
		7/11,
		12/12,
		5/13,
		11/14
		}{
		\node[] (n) at (\i/1.5,\j/3){};
		\node[] (c) at (-0.2,\j/3){};

		\ifthenelse{ \j = 0 \OR \j = 1 \OR \j = 3 }{
			\fill[red](c) circle[radius=1mm];
			}{
			\fill(c) circle[radius=1mm];
		}
		\ifthenelse{ 
		\i = 1 \OR \i = 2 \OR \i = 9
			}{ 
			\fill[red](n) circle[radius=1mm];
			}{
			\fill(n) circle[radius=1mm];
		}
	}
	\node[] at (-0.2,-0.5) {$C$};
	\foreach \i in { 1,2,...,14 }{
		\node[] at (\i/1.5,-0.5) {$C_{\i}$};
	}
\end{tikzpicture}
	\caption{
		This is an example case that each $C_i$ contains only one index.
		The total number of types is the same as that of indices.
		Therefore, one index exclusively corresponds to one type (one-to-one correspondence).
	}
	\label{fig:alldiftypes}
\end{figure*}
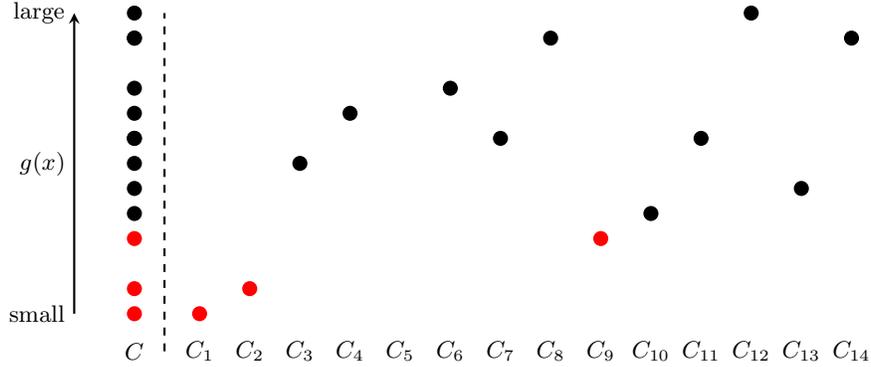

As it treats types, we consider the following conditions (see also Figure~\ref{fig:newcondition}).
\begin{itemize}
	\item Each index has a type.
	\item Let $C_i$ be a set of indices that have type $i$.
	\item Let $c$ be the number of types.
\end{itemize}
D\"urr and H{\o}yer's algorithm finds the indices of the $c$ smallest values each of which is the minimum in each type (see Figure~\ref{fig:types}).
This means that so as to use the conventional algorithm for the general \textit{finding $k$-minima} problem,
all indices must be of different types like Figure~\ref{fig:alldiftypes},
which corresponds to $c=N=|D|$.
Hereafter, we assume all indices are of different types.

Intuitive explanation of the algorithm is that multiple thresholds are maintained while a single threshold is maintained in FM algorithm.
Let $T$ be a set of $k$ thresholds and let $f_T(x)$ be an oracle function such that

\begin{align}
    \label{eqn:FkM_oracle}
	f_T(x)
	=
	\begin{cases}
		1,&
		\textrm{if } g(x)<g(t) \textrm{ for some } t \in T,
		\\
		0,&
		\textrm{if } g(x)\ge g(t) \textrm{ for some } t \in T.
		\\
	\end{cases}
\end{align}
Similar to Eq.~\eqref{eqn:FM_oracle} of FM algorithm, Eq.~\eqref{eqn:FkM_oracle} is regarded as AA for \textit{some} threshold $t$.
However, the meaning of \textit{some} is not clearly mentioned in \cite{durr2006quantum}.
Though the algorithm does not work well in the worst case, which selects the threshold index $t$ that minimizes $g(t)$,
it works in the following cases.
\begin{enumerate}
	\item \label{enu:1}
	$t$ is randomly chosen from $T$.
	\item \label{enu:2}
	$t$ is selected so as to maximize $g(t)$.
\end{enumerate}
In the case of \ref{enu:2}, which is the best case, such $t$ is obtained by \textit{finding maximum} algorithm~\cite{ahuja1999quantum}. 
Hence, its computational burden is $\mathcal{O}(\sqrt{k})$.
It can be ignored, as it is small enough compared to the complexity of the whole algorithm (i.e., $\mathcal{O}(\sqrt{kN})$).

In addition to that, we have to keep the elements of $T$ without duplication.
So as to do that, the algorithm requires duplication check 
or removing $T$ from search indices set
\begin{align}
	M &= \lbrace x \mid f_T (x) =1 , x \in D \rbrace.
\end{align}
We assume that it removes $T$ from $M$ because our proposed method also removes already found indices from $M$.

By ignoring types of data, the \textit{finding $k$-minima} algorithm based on D\"urr and H{\o}yer's algorithm is given as follows.
\begin{enumerate}
	\item Initialize set $T$ as randomly chosen $k$ indices from $D$.
	\item Repeat the following forever.
	\begin{enumerate}
		\item Randomly select a threshold index $t$ from $T$.
		\item Find index $x$ such that $f_T(x)=1$.
		\item Find $t_\textrm{max}$ such that $t_\textrm{max} = \argmax_{t \in T} g(t)$ \footnote{Though this process is not described in \cite{durr2006quantum}, we add this because we think it is required for conversion.}.
		\item Replace threshold index $t_\textrm{max}$ with $x$.
	\end{enumerate}
\end{enumerate}

In this algorithm, $T$ is updated in a step-by-step manner and
each updating step replaces the index that has the maximum value in $T$ with the found index by AA.
This is a kind of a greedy algorithm.

\section{Proposed Finding $k$-Minima Algorithm}

We propose a new \textit{finding $k$-minima} algorithm with the complexity of $\mathcal{O}_q(\sqrt{kN})$.
Our idea is to search a good threshold (the first phase) and use it for \textit{finding all marked $k$-indices} algorithm (the second phase).
We begin with presenting the second phase.
In the second phase, all $k'$ indices, where $k' \ge k$, are found.
Suppose that threshold index $t_{k'}$ satisfy
\begin{align}
	M &= \lbrace x \mid g(x) < g(t_{k'}), x \in D\rbrace, \\
	|M|&=k'.
\end{align}

In the first phase, in order to find the threshold $t_{k'}$, we use FM algorithm and QC algorithm.
As shown in Sec.~\ref{sec:fm}, in the process of FM algorithm, $\mathcal{O}(\sqrt{N})$ threshold indices are found in the descending order of values.
Therefore, it is easy to find $t_{k'}$ from them by a binary search with QC algorithm.

We present more detail about this binary search with QC algorithm.
Let us define the following.
\begin{itemize}
	\item Let $t^{FM}_i$ be the threshold index that is found in the $i$-th step of FM algorithm.
	\item Let $T_{FM}$ be a set of thresholds that are found in the process of FM algorithm, which is given by
	\[
		T_{FM}  = \lbrace t^{FM}_1, t^{FM}_2, \hdots \rbrace
	\]
\item Let $M^{FM}_i$ be a set of marked indices of the $i$-th step in FM algorithm.
\item Let $h(t)$ be a function that maps index $t$ to the number of indices whose values are less than the value of threshold $t$ in $D$, which is counted by QC algorithm.
That is,
\begin{align}
	h(t) &= |M(t)|,
\end{align}
where
\begin{align}
	M(t) &= \lbrace x \mid g(x)<g(t), x \in D \rbrace.
\end{align}

\end{itemize}

The goal of the binary search is to find $i$ such that $h(t^{FM}_{i+1}) \le k < h(t^{FM}_i)$.
Once such $i$ is found, $h(t^{FM}_i)$ is used as $k'$.
Fortunately, $t_{k'}$ exists in the last $k$ indices of all found thresholds.
Hence, we do not have to search all found $\mathcal{O}(\sqrt{N})$ indices but the last $k$ indices.
As QC algorithm requires $\mathcal{O}(\sqrt{N})$ query complexity for $N$ indices~\cite{brassard1998quantum},
a binary search for $k$ indices requires $\mathcal{O}(\log{k})$ comparison.
As QC algorithm has to run in each step of the binary search, threshold $t_{k'}$ can be found in $\mathcal{O}(\sqrt{N}\log{k})$.

In summary, our threshold searching algorithm is shown below.
\begin{enumerate}
	\item Apply FM algorithm and save the indices of the lastly found $k$ thresholds. 
	\item Apply the binary search on the $k$ indices of thresholds.
\end{enumerate}

Once we find such a threshold, we can find all marked $k'$ indices by applying \textit{searching all marked $k$-indices} algorithm.
This algorithm searches all elements in the set $\lbrace x \mid x \in D, f(x)=1 \rbrace$.
For simplicity, we assume that $\mathcal{O}(k')=\mathcal{O}(k)$.
Let $T$ be a set of already found indices in the step of \textit{searching all marked $k$ indices algorithm} and let $f'_{t_k}(x)$ be an oracle such that
\begin{align}
	f'_{t_k}(x)
	=
	\begin{cases}
		1, &
		\textrm{if } g(x)<g(t_k) \textrm{ and } x \notin T,
		\\
		0, &
		\textrm{otherwise}.
	\end{cases}
\end{align}
Then, searching all marked $k$-indices in $M$ can be done in $\mathcal{O}(\sqrt{kN})$.

The whole algorithm of the proposed method is below.
\begin{enumerate}
	\item Apply FM algorithm to $D$ and save the last $k$ indices of FM algorithm step.
	\item Search threshold index $t_{k'}$ by a binary search on $k$ indices.
	The comparison key is $|M^{FM}|$ that is derived by QC algorithm.
	\item Apply \textit{finding all marked $k$-indices} algorithm with threshold $t_{k'}$.
\end{enumerate}

The total query complexity is given as
\begin{align}
	\mathcal{O} (\sqrt{N}\log{k})+\mathcal{O}(\sqrt{kN})=\mathcal{O}(\sqrt{kN}).
\end{align}

------

\section{Conclusion}

In this paper, we proposed a new finding $k$-minima algorithm and derived its query complexity.
Our algorithm is easier to understand and more elegant than D\"urr's algorithm~\cite{durr2006quantum}.
Finding $k$-minima algorithm can be applied to many kinds of algorithms or applications. 
For example, $k$-nearest neighbor search, $k$-nearest neighbor clustering and classification.

\bibliographystyle{unsrt} 


\appendix

\section{Complexity of Searching All Marked $k$-Indices Algorithm \label{sec:findall}}

The problem to search all marked $k$-indices from $D$ and its query complexity are briefly discussed in~\cite{ambainis2004quantum,ambainis2005new,klauck2007quantum,dorn2010note,aimeur2013quantum}.
Here, we will give the complexity of the problem in an easy-to-understand way.

Let $D$ and $M$ be sets of data indices where $|D|=N$ and marked indices where $|M|=k$, respectively.
Since AA searches one of the $k$ indices from $D$ in $\mathcal{O}_q(\sqrt{N/k})$ query complexity~\cite{boyer1996tight,grover1996fast,brassard2002quantum},
a way to search all marked $k$-indices is to repeat the following process until all $k$-marked indices are found.
\begin{easylist}[enumerate]
	@ Find one index $m$ from $M$ by using AA.
	@ Remove $m$ from $M$.
\end{easylist}

\begin{figure}[t]
	\input{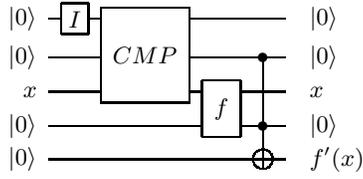}
	\caption{
		An example of quantum circuit $f'$. 
		The gate $T$ creates the superposition of already found indices in $T$.
		The gate $f$ is an original oracle of AA that finds one index from $D$.
		We can use Fredkin gate instead of Toffoli gate because we use this gate as AND gate~\cite{nielsen2002quantum}. 
		As we said before, we omit inversing gates here.
		Strictly speaking, the gate $\text{CMP}$ do not guarantee that the input $x$ and output are the equivalent superposition.
		To avoid this problem we have to add an inverse gate after $\text{CMP}$ gate.
		However, we assume that the inversing gate is already built in the $\text{CMP}$ gate for simplicity.
	}
	\label{fig:saaa}
\end{figure}

By updating $f(x)$ to $f'(x)$ to remove already found indices.
\begin{align}
	f'(x)=
	\begin{cases}
		1, &
		\text{if }
		f(x)=1 \text{ and } x \notin T,
		\\
		0, &
		\text{if }
		f(x)=0 \text{ or } x \in T.
	\end{cases}
\end{align}
This $f'(x)$ needs additional quantum circuit on $f(x)$. However, such quantum circuit is not complex, because we can easily make superposition of the already found indices $T$ in $\mathcal{O}_g(\log{N})$~\cite{grover2002creating,kaye2004quantum,soklakov2006efficient,lloyd2013quantum}.
Figure~\ref{fig:saaa} shows an example of the quantum circuit $f'$.

The following algorithm finds all $k$-indices.
\begin{easylist}[enumerate]
	@ Repeat the following process $\mathcal{O}(k)$ times.
			@@ Search one index $m \in M$ from $D$ by using AA and $f'(x)$.
			@@ Add a found index $m$ to $T$.
			@@ Remove a found index $m$ from $M$ as given as
				\begin{align*}
					M&=M\backslash m.
				\end{align*}
\end{easylist}
In the $t$-th iteration of this algorithm, as $|M|=k-t$, the query complexity to search a marked index is $\mathcal{O}_q(\sqrt{\frac{N}{k-t}})$.
Therefore, the total query complexity is given as
\begin{align}
	&\mathcal{O}_q\left(
		\sqrt{\frac{N}{k}}
		+
		\sqrt{\frac{N}{k-1}}
		+
		\cdots
		+
		\sqrt{N}
	\right)
	\\
	=&
	\label{eqn:search_all_k_indeces_complexity}
	\mathcal{O}_q(\sqrt{kN}),
\end{align}
because
\begin{align}
	\sum_{t=1}^{k}
	\sqrt{\frac{1}{t}}
	<
	1+ \int_{1}^{k}\sqrt{\frac{1}{t}} dt
	= 2\sqrt{k} -1.
\end{align}

%

\end{document}